\documentclass[aps,prb,twocolumn,superscriptaddress,showpacs]{revtex4-1}

\usepackage{graphicx}
\usepackage{calc}
\usepackage{bm}
\usepackage{color}
\usepackage{mathtools}
\usepackage{gensymb}

\begin{document}

\title{Demonstration of  the third-order nonlinear Hall effect in topological  Dirac semimetal $NiTe_2$}

\author{V.D. Esin}
\author{A.V.~Timonina}
\author{N.N.~Kolesnikov}
\author{E.V.~Deviatov}
\affiliation{Institute of Solid State Physics of the Russian Academy of Sciences, Chernogolovka, Moscow District, 2 Academician Ossipyan str., 142432 Russia}

\date{\today}

\begin{abstract}
We experimentally investigate  third-order nonlinear Hall effect for three-dimensional $NiTe_2$ single crystal samples. $NiTe_2$ is the recently discovered type-II Dirac semimetal, so both the inversion and the time-reversal symmetries are conserved in the bulk. As a result, the well known second-order nonlinear Hall effect does not expected for this material, which we confirm as negligibly small second-harmonic transverse Hall voltage response to the longitudinal  ac electric current. As the main experimental result, we demonstrate the unsaturated third-harmonic  Hall response in $NiTe_2$, which well corresponds to the theoretically predicted third-order nonlinear Hall effect in Dirac semimetals. We also demonstrate, that the third harmonic signal does not depend on the external magnetic field, in contrast to the field-depended first-order and second-order  Hall effects.
\end{abstract}

\maketitle

\section{Introduction}

Topological materials are the powerful platform for nonlinear effects realization. For example, the second-order nonlinear Hall effect (NLHE), which is predicted as a nonzero transverse voltage response at both zero and twice the frequency,  was theoretically studied~\cite{nlhe-t,nlhe-t2,nlhe-t3,nlhe-t4,nlhe-t5} and experimentally demonstrated in a number of topological materials, such as Weyl and Dirac semimetals, chiral semimetals and magnetic nodal-line semimetals~\cite{nlhe-exp,nlhe1,nlhe2,nlhe-exp2,nlhe-exp3}. 

It was argued~\cite{nlhe-t,berry,berry2}, that a nonlinear Hall-like response may occur due to the Berry curvature, which can be understood as a magnetic field in momentum space due to some symmetry requirements. In particular, the second-order NLHE appears in materials with time-reversal symmetry, but it usually requires the breaking of inversion symmetry~\cite{nlhe-t6}. In this case, electric current generates effective sample magnetization due to the Berry curvature dipole, which is proportional to the current. It leads to the appearance of a transverse Hall voltage response $V_{xy}\sim I^2$ even in the case of time-reversal symmetry. The Hall voltage can be measured as the second-harmonic response  $V^{2\omega}_{xy}$ for the a.c. bias current $I$.

However, even the both time-reversal and inversion symmetries are present, the third-order NLHE can still be observed. It arises as a generalization of Berry curvature concept and is a result of the non-zero $\tilde{G}$ Berry connection polarizability tensor~\cite{bloch-3w,bloch-3w-2,berry3w,topis3w,berry-tensor,weyl3w,berry3w-3}. By definition, it is the proposed~\cite{bloch-3w,bloch-3w-2} second-rank tensor, which reflect the relationship between the field-induced Berry connection and the applied electric field $\tilde{G}=$${\partial A(\boldsymbol{k})}\over{\partial \boldsymbol{E}}$, where $A(\boldsymbol{k})$ is the Berry connection, $\boldsymbol{k}$ is wave vector and $\boldsymbol{E}$ is the electric field~\cite{berry3w,topis3w}. 

The Berry connection $A(\boldsymbol{k})$ is independent of electrical field in the case of second-order NLHE~\cite{berry-tensor}. In contrast, for third-order NLHE, the Berry connection $A(\boldsymbol{k})$ is modulated by an electric field $\boldsymbol{E}$ using a Berry connection polarizability tensor $\tilde{G}$, which leads to a field-induced Berry curvature $\boldsymbol{\Omega}^E=\nabla_k\times A(\boldsymbol{k,E})$~\cite{weyl3w,berry3w-3}. The latter appear as the third-harmonic transverse Hall voltage response V$^{3\omega}_{xy}$.

Dirac semimetals are a well-studied type of topological materials at present time~\cite{armitage}. It's peculiar properties are connected with the  energy band  inversion. Particularly, Dirac semimetals are characterized by gapless spectrum due to band touching in some distinct points, which are the special points of Brillouin zone. 

For an ideal Dirac semimetal, both the inversion and the time-reversal symmetries are conserved in the bulk, which should lead to the disappearance of the second-order NLHE response. At the same time,  third-order NLHE has already been observed~\cite{dirac3w} for the typical  Dirac semimetal Cd$_3$As$_2$. Since this effect has the topological origin,  third-order NLHE should be demonstrated to be independent of the particular material.
On the other hand, NLHE can be used for high-frequency (even terahertz or infrared) detection~\cite{NLHErect,NLHErect1}, which is important for wide-band communications~\cite{tera},  wireless charging, energy harvesting, etc. An advantage of the NLH rectification is the absence of thermal losses, since it originates from the Berry curvature dipole. The latter can be in principle controlled by electric field, which has been demonstrated~\cite{NLHEgate} for two-dimensional WTe$_2$. Significant rectification has been also predicted in centrosymmetric metals via third-order nonlinear response~\cite{rect3w}, therefore, it is reasonable to investigate third-order NLHE for different Dirac materials.

One of the promising candidates for the third-order nonlinear Hall effect investigations  is the $NiTe_2$, which has been recently identified as type-II Dirac semimetal. The Dirac spectrum has been experimentally confirmed for $NiTe_2$ by angle-resolved photoemission spectroscopy (ARPES)~\cite{arpes,arpes2}. 

Here,  we experimentally investigate  third-order nonlinear Hall effect for three-dimensional $NiTe_2$ single crystal samples. $NiTe_2$ is the recently discovered type-II Dirac semimetal, so both the inversion and the time-reversal symmetries are conserved in the bulk. As a result, the well known second-order nonlinear Hall effect does not expected for this material, which we confirm as negligibly small second-harmonic transverse Hall voltage response to the longitudinal  ac electric current. As the main experimental result, we demonstrate the unsaturated third-harmonic  Hall response in $NiTe_2$, which well corresponds to the theoretically predicted third-order nonlinear Hall effect in Dirac semimetals. We also demonstrate, that the third harmonic signal does not depend on the external magnetic field, in contrast to the field-depended first-order and second-order  Hall effects.

\section{Samples and technique}

\begin{figure}
\includegraphics[width=0.75\columnwidth]{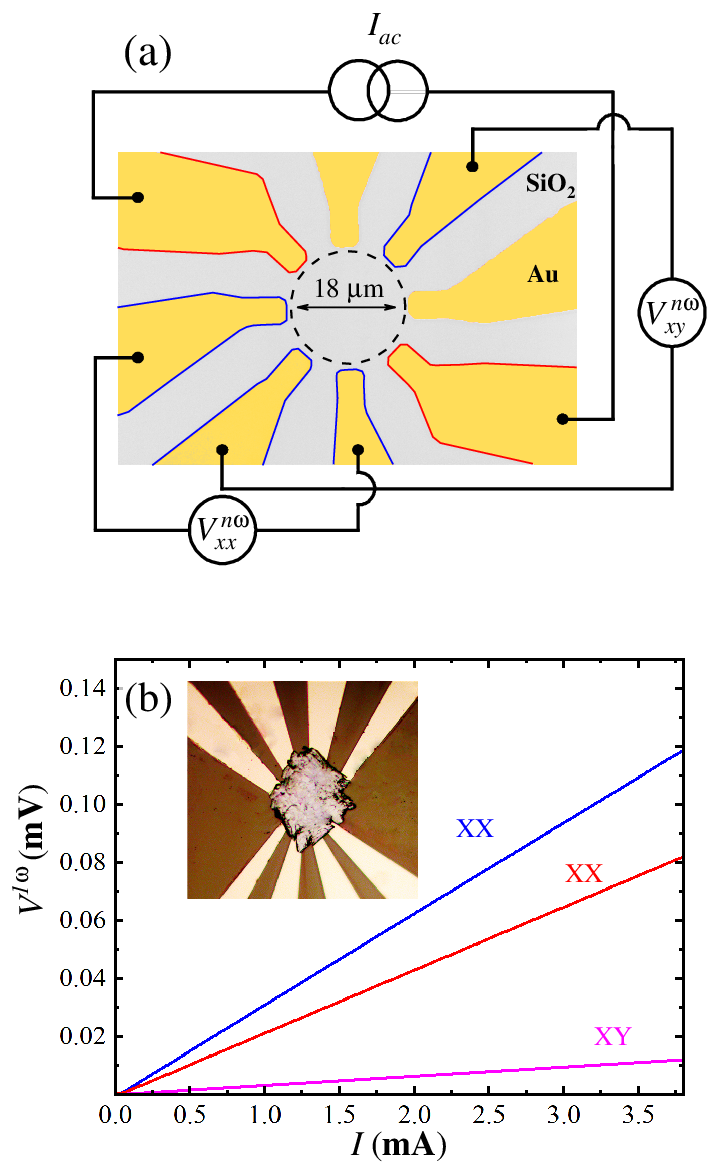}
\caption{(Color online) (a) Sketch of the sample with the electrical connections. 100 nm thick Au leads are formed by lift-off on the $SiO_2$ substrate.  A 0.5 $\mu$m thick $NiTe_2$ single crystal  is placed on top of the lead. We study electron transport by the standard four-point lock-in technique: two contacts (highlighted in red) are used for the current and ground. Other contacts (highlighted in blue) are  for V$^{n\omega}_{xx}$ and V$^{n\omega}_{xy}$ (where n = \{1,2,3\}) voltage components measurements.  (b) Examples of low-temperature longitudinal V$^{1\omega}_{xx}$ (red and blue lines for two samples, respectively) and transverse V$^{1\omega}_{xy}$ (magenta line) first-harmonic voltage  components in zero magnetic field. Strictly Ohmic dependences of the V$^{n\omega}_{xx}$ signal and much smaller V$^{1\omega}_{xy}$ confirm correctness of th experimental geometry. Inset shows the optical image of the sample.}
\label{fig1}
\end{figure}

$NiTe_2$ was synthesized from elements, which were taken in the form of foil (Ni) and pellets (Te). The mixture was heated in an evacuated silica ampule up to 815$\degree$ C with the rate of 20 deg/h, the ampule was kept at this temperature for 48 h. The crystal was grown in the same ampule by the gradient freezing technique with the cooling rate of 10 deg/h. As a result, we obtain 80 mm long and 5 mm thick $NiTe_2$ single crystal, with (0001) cleavage plane.

The powder X-ray diffraction analysis (Cu $K\alpha$1 radiation, $\alpha$ = 1.540598 $\dot{A}$) confirms single-phase $NiTe_2$ with P-3m1 (164) space group (a = b = 3.8791 $\dot{A}$, c = 5.3005 $\dot{A}$). The known structure model is also refined with single crystal X-ray diffraction measurements (Oxford diffraction Gemini-A, Mo K$\alpha$). Nearly stoichiometric ratio Ni$_{1-x}$Te$_2$ ($x < 0.06$) is verified by the energy-dispersive X-ray spectroscopy.

The quality of our $NiTe_2$ material was tested in standard four-point magnetoresistance measurements~\cite{qual1}. In particular, non-saturating longitudinal magnetoresistance~\cite{qual2,qual3} is shown for our $NiTe_2$ samples in normal magnetic field~\cite{qual1}. Despite the $NiTe_2$ can be thinned down to two-dimensional monolayers, topological materials are essentially three-dimensional objects~\cite{armitage}. We select relatively thick (above 0.5 $\mu m$) $NiTe_2$ single crystal flakes for our samples. Additionally, thick samples also ensures their homogeneity, which is important for reliable xx- and xy- resistance separation. 

In order to create samples with thick flakes we use special technique, which is known to provide high quality Ohmic contacts~\cite{tech1,tech2,tech3}. Fig.~\ref{fig1} (a) shows a sketch of a sample. The leads pattern is formed by thermal evaporation of 100 nm Au on a oxidized SiO$_2$ silicon substrate and subsequent lift-off technique. To obtain  the xx- and -xy- components, the pattern is of circular geometry, as depicted in Fig.~\ref{fig1} (a). The distance between two neighbor contacts is 5 $\mu$m, while the diameter of the circle is 18 $\mu$m. As a second step, we select the fresh mechanically exfoliated $NiTe_2$ flake and transfer it to the center of the leads pattern. Finally, the flake is slightly pressed by another oxidized silicon substrate. Afterward,  the flake is fixed to the Au leads and does not require further external pressure.As a result, the ohmic Au-$NiTe_2$ junction are formed on the SiO$_2$ substrate. This procedure provides transparent junctions, stable in different cooling cycles as it has bee verified for a wide range of topological materials~\cite{tech1,tech2,tech3,qual1,cosns,andreev,mnte}. The optical image of the  sample is shown in the inset to Fig.~\ref{fig1} (b).

We measure both the V$^{n\omega}_{xy}$ and V$^{n\omega}_{xx}$ (where n = \{1,2,3\}) voltage response components by standard four-point lock-in technique. The electrical scheme is shown in Fig.~\ref{fig1} (a): two contacts are used for ac current $I$ and ground (highlighted in red line), while other contacts (highlighted in blue line) are for V$^{n\omega}_{xx}$ and V$^{n\omega}_{xy}$ measurements. 

In principle, nonlinear voltage response can arise due to thermoelectricity, since topological materials are usually characterized by strong thermoelectric effects~\cite{thermo1,thermo2}. To exclude this possible contribution, we choose strictly symmetric contact configurations without  temperature gradients between the potential probes. Also, we check that the measured voltage is antisymmetric with respect to the voltage probes swap. The lock-in signal is independent of the modulation frequency in wide range 0.1 kHz - 1 kHz due to the filters applied. All the measurements are performed in a standard 1.4 K – 4.2 K cryostat equipped with superconducting solenoid.

 \begin{figure}
\includegraphics[width=\columnwidth]{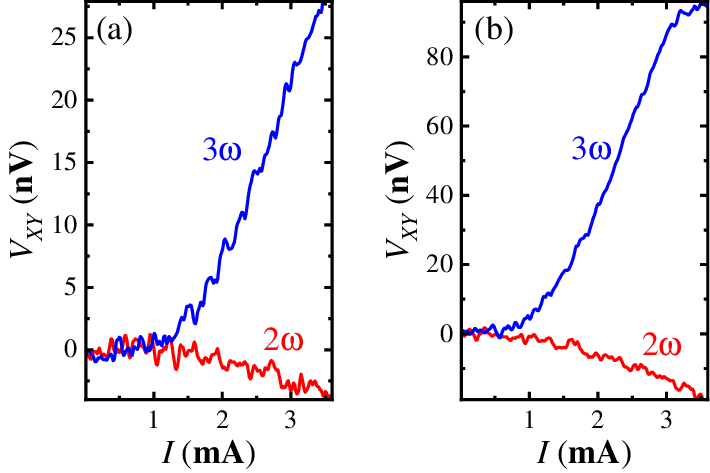}
\caption{(Color online) Examples of the transverse $V_{xy}$ voltage components in dependence on the ac current amplitude $I$ in zero magnetic field. The data are shown for two different samples, (a) and (b) panels, respectively.  Figure demonstrates  finite non-saturating V$^{3\omega}_{xy}$ third-harmonic Hall voltage (blue curves). In contrast, the second-harmonic $V^{2\omega}_{xy}$ is an order of magnitude smaller (red curves)  for both samples. This behavior is confirmed for different cooling cycles and is temperature-independent within the 1.4-4.2 K range. 
 }
\label{fig2}
\end{figure}

\section{Experimental results}

In order to confirm the correctness of the experimental geometry we measure standard first-harmonic  voltage response, see  Fig.~\ref{fig1} (b). By increasing ac current amplitude $I$, we demonstrate well-developed Ohmic behavior for the longitudinal V$^{1\omega}_{xx}$ voltage component in zero magnetic field for two different samples (blue and red curves, respectively). The slopes of V$^{1\omega}_{xx}$ dependence correspond to $\approx 0.02- 0.03\ohm$ bulk resistance. In contrast, the typical transverse V$^{1\omega}_{xy}$ voltage is an order of magnitude smaller than the V$^{1\omega}_{xx}$ one (the magenta curve in Fig.~\ref{fig1} (b)) in zero magnetic field. Also, 
we verify  the linear dependence of V$^{1\omega}_{xy}$ on the magnetic field at fixed ac current, as it should be expected for the conventional Hall effect to finally ensure the correctness of the  experimental geometry.

Fig.~\ref{fig2} shows  typical examples of the transverse $V_{xy}$ voltage components for two different samples ((a) and (b) panels, respectively) in zero magnetic field. For both samples, we observe a well-developed nonlinear $V^{3\omega}_{xy}$ third-harmonic voltage response, see blue curves in Fig.~\ref{fig2}. The absolute value of the signal is from 30 to 100 $nV$ for 3.85 $mA$ maximal current for these two samples. This behavior is confirmed for different cooling cycles and is temperature-independent within the 1.4-4.2 K range. 

\begin{figure}
\includegraphics[width=\columnwidth]{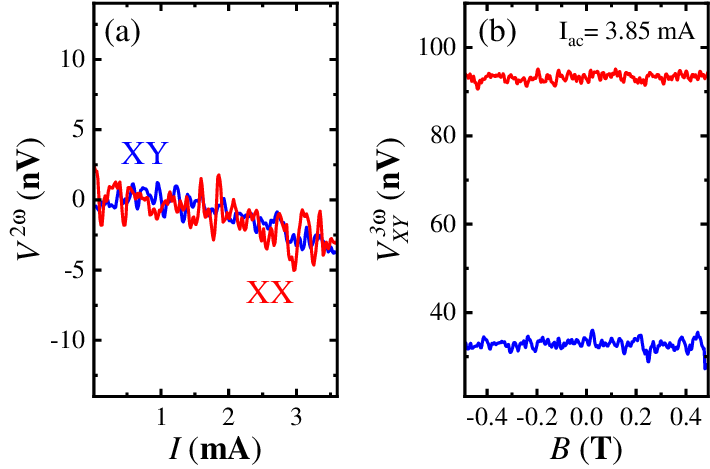}
\caption{(Color online)  (a) The second-harmonic $V^{2\omega}_{xy}$ and $V^{2\omega}_{xx}$ voltage components in zero magnetic field. Both curves coincide well within  the 3.85 mA ac current range, which seems to reflect the accuracy of the experimental geometry. (b) The third-harmonic voltage $V^{2\omega}_{xy}$ signal dependence on the magnetic field $B$ at fixed ac current I$_{ac}$=3.85 mA for two different samples (blue and red curves, respectively). The absolute value of $V^{3\omega}_{xy}$ signal can vary from sample to sample ($\approx$30~nV and $\approx$90~nV for the blue and the red curves, respectively), while there is no dependence of the $V^{3\omega}_{xy}$ signal on the magnetic field, in contrast to the prominent dependence for the second-harmonic NLHE in Weyl semimetals~\cite{nlhe1,nlhe-exp3,mandal,zyuzin}
 }
\label{fig3}
\end{figure}

The third-order NLHE is expected in Dirac semimetals with both time-reversal and inversion symmetries. In this case, one can expect no second-harmonic transverse voltage response~\cite{nlhe-t,nlhe-t2,nlhe-t3,nlhe-t4,nlhe-t5}.  In Fig.~\ref{fig2}, the measured values of the $V^{2\omega}_{xy}$ signal is from 5 to 20 $nV$(red curves, which is an order of magnitude smaller than for third-harmonic $V^{3\omega}_{xy}$ component. Moreover, Fig.~\ref{fig3} (a) demonstrates that both
the longitudinal $V^{2\omega}_{xx}$ second-harmonic signal and  the transverse $V^{2\omega}_{xy}$  coincide well within  the 3.85 mA ac current range, which seems to reflect the accuracy of the experimental geometry. 

In contrast to the first-order and second-order  Hall effects~\cite{nlhe1,nlhe-exp3}, the external magnetic field has no effect on the  $V^{3\omega}_{xy}$ third-harmonic transverse voltage in Fig.~\ref{fig3} (b). For two different samples (blue and red curves, respectively), we sweep magnetic field within the $\pm$0.5~T range at the fixed ac current amplitude $I = 3.85$ $mA$. The absolute value of the $V^{3\omega}_{xy}$ signal can vary from sample to sample ($\approx$30~nV and $\approx$90~nV for the blue and the red curves, respectively), while there is no dependence of the $V^{3\omega}_{xy}$ signal on the magnetic field. 

\section{Discussion} \label{disc}

As a result, we demonstrate the third-order NLHE as the third-harmonic  transverse Hall voltage response $V^{3\omega}_{xy}$ for Dirac semimetal  $NiTe_2$. This conclusion is confirmed by almost zero value of the second harmonic $V^{2\omega}_{xy}$ voltage response, in good correspondence with theoretical predictions~\cite{berry-tensor,berry3w,berry3w-2,weyl3w,topis3w,bloch-3w,berry3w-3}. 

Despite the   third-order NLHE has already been observed~\cite{dirac3w,weyl3w2,ferro3w,ferro3w2} for some free- and two-dimensional materials,  especially for the typical  Dirac semimetal~\cite{dirac3w} Cd$_3$As$_2$, our experiment confirms its independence of the particular material. In other words,  it confirms  the topological origin of the third-order NLHE.

Thermoelectric effects may be a reason of the nonlinear response at the higher harmonic. For example, Joule heating is proportional to the  square of the current, so it can produce significant $V^{2\omega}_{xy}$ or $V^{2\omega}_{xx}$ voltage components in the case of asymmetric voltage probes.  However,  we show almost zero both transverse $V^{2\omega}_{xy}$ and longitudinal $V^{2\omega}_{xx}$ voltage components, which is a good argument of the correct geometry of our experiment. Moreover, the third-harmonic voltage response is an order of magnitude higher in Fig.~\ref{fig2}, which also excludes  Joule heating effects. There are less significant reasons, leading to contribution on nonlinear transport~\cite{weyl3w2}, e.g. the capacitance coupling effect. Higher order voltage response can also be caused by a "parasitic" capacitance connected inside the circuit. However, we check that our transverse $V^{n\omega}_{xy}$ signal does not depend on the frequency in  a wide range 0.1 kHz - 1 kHz.

Another possible reason for $V^{3\omega}_{xy}$ response is side jump and skew scattering on the magnetic impurities, which can lead both transverse and longitudinal nonlinear responses~\cite{berry3w,skew,skew1}. However, similar contribution can only appear in  noncentrosymmetric crystals, while the Ref.~\cite{nite2str} insists that the $NiTe_2$ single crystal has the centrosymmetric structure. In contrast, it is argued in Refs.~\cite{berry-tensor,berry3w,berry3w-2,weyl3w,topis3w,bloch-3w,berry3w-3}, that a third-order Hall-like voltage response can be caused by Berry connection polarizability tensor $\tilde{G}$ in centrosymmetric materials.  

As a new and important observation, the external magnetic field has no effect on the  $V^{3\omega}_{xy}$ third-harmonic transverse voltage in Fig.~\ref{fig3} (b), despite  the first-order and second-order  Hall effects depend on the magnetic field.  In the  former case the dependence is the classical Hall effect, while $B$-like correction is also predicted for the second-harmonic NLHE~\cite{mandal,zyuzin}. In experiments,  field-dependent $V^{2\omega}_{xy}(B)$ has been demonstrated for different Weyl semimetals~\cite{nlhe1,nlhe-exp3}. 

On the other hand, applying of  magnetic field  should not lead to the disappearance of the third-order NLHE response. The third-order NLHE has been observed~\cite{ferro3w2} in  two-dimensional ferromagnetic $Fe_5GeTe_2$. Moreover, Ref.~\cite{berry-tensor} insists that there are magnetic symmetry groups in which a third-order NLHE is possible. Despite further theoretical efforts are necessary, the $V^{3\omega}_{xy}$ independence of the magnetic field allows to firmly distinguish the third-order NLHE in Dirac semimetal from the well-known  first-order and second-order  Hall effects.

\section{Conclusion}

As a conclusion, we experimentally investigate  third-order nonlinear Hall effect for three-dimensional $NiTe_2$ single crystal samples. $NiTe_2$ is the recently discovered type-II Dirac semimetal, so both the inversion and the time-reversal symmetries are conserved in the bulk. As a result, the well known second-order nonlinear Hall effect does not expected for this material, which we confirm as negligibly small second-harmonic transverse Hall voltage response to the longitudinal  ac electric current. As the main experimental result, we demonstrate the unsaturated third-harmonic  Hall response in $NiTe_2$, which well corresponds to the theoretically predicted third-order nonlinear Hall effect in Dirac semimetals. We also demonstrate, that the third harmonic signal does not depend on the external magnetic field, in contrast to the field-depended first-order and second-order  Hall effects.

\acknowledgments

We wish to thank  S.S~Khasanov for X-ray sample characterization.  We gratefully acknowledge financial support  by the  RF State task.

\end{document}